\documentclass[aps,pre,showpacs,twocolumn,english]{revtex4}
\pdfoutput=1
\usepackage{bm}
\usepackage{graphicx}
\bibstyle{approve.bib}
\usepackage{setspace}
\usepackage[]{fontenc}
\usepackage[latin1]{inputenc}
\usepackage{prettyref}

\makeatletter

\usepackage{amsthm,amssymb}
  \theoremstyle{plain}

\newrefformat{fig}{Figure~\ref{#1}}
\newrefformat{tab}{Table~\ref{#1}}

\usepackage{babel}
\makeatother

\begin{document}
\title{Compositionality, stochasticity and cooperativity in dynamic 
models of gene regulation}
\author{Ralf Blossey}
\affiliation{Biological Nanosystems, 
Interdisciplinary Research Institute c/o IEMN, Cit\'e
Scientifique BP 60069, F-59652 Villeneuve d'Ascq, France}
\author{Luca Cardelli}
\affiliation{Microsoft Research, Roger Needham Building, 7 J.J. Thomson
Avenue, Cambridge CB3 0FB UK}
\author{Andrew Phillips}
\affiliation{Microsoft Research, Roger Needham Building, 7 J.J. Thomson
Avenue, Cambridge CB3 0FB UK}
\date{\today}

\begin{abstract}
We present an approach for constructing dynamic models for the simulation
of gene regulatory 
networks from simple computational elements. Each element is called a ``gene 
gate'' and defines an input/output-relationship corresponding to the binding 
and production of transcription factors. The proposed reaction kinetics of the 
gene gates can be mapped onto stochastic processes and the standard ode-description.
While the ode-approach requires fixing the system's 
topology before its correct implementation, expressing them in stochastic 
$\pi$-calculus leads to a fully compositional scheme: network elements become 
autonomous and only the input/output relationships fix their wiring. The modularity
of our approach allows to pass easily from a basic first-level description to 
refined models which capture more details of the biological system.
As an illustrative application we present the stochastic repressilator, an 
artificial cellular clock, which oscillates readily without any cooperative effects. 
\end{abstract}


\maketitle

\section{Introduction}

Providing efficient ways to model the dynamics of gene regulatory networks
is an important challenge in systems biology. Many different methods have 
been proposed in the past for such dynamical networks. One prominent
approach is based on discrete logical methods, going back to the pioneering 
work by Kauffman on synchronous Boolean networks (Kauffman S A, 1969) and 
Thomas on asynchronous Boolean networks (Thomas R, 1973 $\&$ 1991); reviews of
the current state of such approaches are (de Jong H, 2002, Smolen P et al., 
2002). A different, independent approach is based on 
rate equations, hence on the continuous dynamics of nonlinear ODE's 
(Goldbeter A, 1996). Finally, there are various variants of stochastic
methods, based either on the master equation approach (Van Kampen N G, 1992) 
or on the equivalent Gillespie algorithm (Gillespie D, 1977).

The basic underlying problem for the quantitative description of the dynamics 
of gene regulatory networks is the enormous diversity of the `actors' involved, i.e., 
the biomolecules which determine the network structure and dynamics. 
Both from an analytic and a computational point of view, one therefore needs to 
simplify in order to make simulations of such networks feasible: 
representing all actors by individual computational elements is 
simply unfeasible. But this is not the only problem. Two obvious other
challenges are: i) to have flexible modeling schemes, and ii) schemes 
which do not grow too fast with the increase of the number of reactions 
included. 

In this paper we propose an approach to models of gene regulatory network dynamics 
which is both flexible and has such advantages in terms of system size. 
It combines two features: 

Firstly, our modeling approach to gene regulatory networks is based on an abstraction of 
the genome as a set of input-output elements, the {\it gene gates} (Blossey R et al, 2006). 
The properties of each gate are defined by a set of abstract kinetic reactions. 
(In the simplest - Boolean - setting, a gate would be either on or off.) 
Based on these modules, regulatory circuits can be constructed by formulating input-output relationships between the gates. 
An advantage of this modeling approach is that it allows to start with a very simple 
construction of the gates to represent the overall topology of the network. We show
how more biological detail can be added to the model while leaving the underlying 
topology of the network unaltered. The approach therefore permits to build
computational models with variable degrees of detail which is highly desirable
given the incomplete knowledge of most biological systems.

Secondly, the full advantage of our {\it compositional} approach can 
be seen by formulating the networks in terms of processes defined in a 
process calculus, the $\pi$-calculus, which originates in the field of 
programming languages in theoretical computer science (Milner R, 1999),
and has been proposed for applications to systems biology only recently 
(Priami C et al, 2001, Regev A $\&$ Shapiro E, 2002, Regev A, 2003). 
Not only do the compositional features of this calculus allow to express 
each gate as an autonomous network element, they also significantly reduce 
the system size (Cardelli L, 2007). For $2n$ elements, the size of the 
input/output-interface equals $2n$, while
the number of kinetic reactions can be in the worst case $n^2$.
An introduction to the stochastic $\pi$-calculus and its use in simulations 
is presented in the Supplementary Material (see also Phillips A $\&$ Cardelli L, 2007). 

The process calculus directly allows for a stochastic formulation 
of the dynamics, which is clearly more realistic for networks of molecules 
with small copy numbers than the deterministic dynamics. This feature can 
indeed be critical for the description of the network dynamics. 

We illustrate this by our application of the approach to the repressilator, a 
three-gate inhibitory network which is an artificial cellular clock 
realised experimentally (Elowitz M E and Leibler S, 2000). While
the repressilator readily oscillates within a stochastic dynamics without
cooperative mechanisms in the interaction between genes and transcription factors, 
such cooperativity is required to bring about oscillations in a deterministic (nonlinear) 
gate dynamics, as is shown in Section IV where we map the gene gate reaction
kinetics onto deterministic ode's.

Finally, we demonstrate the refinement of  the basic description to include more 
details of the biological system for the repressilator by an inclusion of the 
transcription, translation and repressor binding processes. Protein complexation 
is found to regularize the oscillations.

\section{Modeling gene regulatory networks by gene gates}

\subsection{The definition of a gene gate}

To be specific, in this work we want to consider genetic interactions 
in genomes similar to those of prokaryotes (bacteria). In such organisms, 
the basic regulatory mechanism follows the classical dogma of molecular 
biology, according to which DNA ``makes" RNA which in turns ``makes" 
protein (Alberts B et al, 2002). 

The modeling scheme we propose for the gene regulatory circuits of such organisms
is based on the idea that the action of 
each gene is uniquely identifiable by its regulatory input (activation/inhibition) 
and its regulatory output. In a first modeling step this therefore amounts to neglect 
all intermediate steps, which are the formation of the gene-transcription factor complex, the recruitment of the polymerase for basal transcription, the transcription process of the gene, the translation process 
of the mRNA. Only the blocking of the gene by transcription factor recruitment (since in the 
present paper we only discuss repressed genes) and the production of a corresponding 
transcription factor from the gene is retained. 

We represent the whole gene network as a composition of ``gene gates". 
A gene gate comprises not only all of the processes listed above but
in addition also the degradation machinery of the proteins. In a gene gate, transcription 
and translation are lumped together in one parameter set, and protein degradation will 
be controlled by a seperate parameter. 

The physical basis for this initial modeling employing a reduction of variables is based on 
the common distinction between slow and fast variables. The selection of these variables is 
indeed important, as has been discussed in detail e.g. in (Bundschuh R et al, 2003).
The advantage of our compositional/modular approach is that we can add all neglected 
intermediate layers of regulation in an easy fashion without 
affecting the basic topology of the network. In this way, the faster processes that were 
neglected in the beginning can be added in principle without any further approximations,
as we demonstrate here.

We thus arrive at the representation of a gene regulator element
as shown schematically in Figure 1 (top left). The double helix represents an 
active gene $g$, 
while the double helix with a blocked promoter region represents a blocked 
gene $g'$. The red and orange ovals represent different types of proteins, 
while the dotted ovals represent degraded proteins. The shapes are labelled 
with a gene name ($g, g'$) or a protein name ($A,B$).

The graphical representation has a precise correspondence with the 
gate reaction kinetics shown at the bottom of the Figure, which summarises
the possible reactions between the gate and the proteins. These reactions
are the blocking of the gene by protein $A$, the production of protein $B$
by the unblocked gene, the unblocking of $g'$, and the decay of the protein,
all with their corresponding rates. 

\subsection{Gene gates in stochastic $\pi$-calculus}

The stochastic $\pi$-calculus is essentially a modular language for describing
the dynamics of a biological system, from which a set of reaction
equations can subsequently be derived. The stochastic $\pi$-calculus
differs from reaction equations in two fundamental ways. Firstly,
instead of modeling the individual reactions of a system we model
its components. This allows a system to be described in a modular
fashion, so that each component can be modified independently. Secondly,
instead of explicitly saying which component can interact with which
other component, we describe the different sites on which a component
can interact. This adds a layer of abstraction to the model, where
two components can interact if they have complementary sites.

Figure 1 (left and right) compares the reaction
equation model and a stochastic $\pi$-calculus model of the gene gate. As
for the reaction kinetics, the graphical representations at the top of the figure 
are equivalent to the textual representations at the bottom. Each shape in the 
graphs represents a protein or gene in a particular state. 

For the stochastic $\pi$-calculus mode on the right, each labelled edge represents a 
reaction, which can be either unary or binary. Unary reactions are 
labelled with a reaction name, where each name is associated with a 
corresponding rate. For example, a protein can degrade by doing a reaction 
$\delta$, and a blocked gene can unblock by doing a reaction $\eta$. A gene 
can also produce a new protein in parallel with itself by doing a 
reaction $\varepsilon$, where a horizontal bar represents parallel composition. 
Binary reactions are labelled with a reaction site preceded by a send 
(?) or receive (!), where each site is associated with a corresponding 
rate. For example, a gene can become blocked by receiving on site $a$, and 
a protein can react by sending on site $b$. Two entities can interact by 
sending and receiving on the same site, where the rate of the reaction 
is equal to the rate of the site. As a result, a protein that sends on 
site a can interact with a gene that receives on $ a$, causing the gene to 
block.

Each shape in the model is parameterised by its interaction
sites. The genes $g,g'$ are parameterised by sites $a$ and $b$,
while the protein $P$ is parameterised by site $b$. Thus $g(a,b)$
denotes a gene that receives on $a$ and that produces proteins which
send on $b$. The parameters allow networks of arbitrary complexity
do be constructed from a single model of a gene gate. For instance,
an autoinhibitory gate can be defined as $g(a,a)$, i.e. a gene that
receives on $a$ and that produces proteins which send on $a$. 
A bistable network can be defined as $g(a,b)\mid g(b,a)$ and a repressilator
network can be defined as $g(a,b)\mid g(b,c)\mid g(c,a)$. 

If we compare the two models in Figure 1
we observe that the reaction equation model contains two proteins
$A$,$B$, but does not fully describe the behaviour of either. In
particular, there is no information on how protein $A$ is produced
or degraded, or on how protein $B$ interacts. In contrast, the stochastic
$\pi$-calculus model describes the complete behaviour of the protein
$P$ that is produced by the gene. Furthermore, the model does not
need to explicitly mention protein $A$, since it only considers the
site on which the gene can interact. This ability to describe the
components of a system in a modular way is one of the main advantages
of the stochastic $\pi$-calculus. Not only does this allow for more maintainable
models, but it can also help to significantly reduce the model size.
Consider the gene network described in Figure 2,
consisting of $N$ proteins $P_{1},\ldots,P_{N}$, each of which can
block $M$ genes $g_{1},\ldots,g_{M}$. For the reaction equation
model we need to explicitly state which protein can block which gene,
resulting in a model of size $N\times M$. In contrast, for the stochastic
$\pi$-calculus model we only need to state that each protein can send
on site $a$ and that each gene can receive on $a$, resulting in
a significantly smaller model of size $N+M$.

\section{Application: The repressilator in stochastic $\Pi$-calculus}

\subsection{Parameter Variation of a Basic Repressilator\label{sub:Parameter-Variation}}

In the first instance we explore the parameter space of a simple repressilator
network, constructed using the gene gate described in Figure 1.
Our compositional approach to modeling allows the network to be defined
in a straightforward manner as $g(a,b)\mid g(b,c)\mid g(c,a)$. Note
that the initial population of proteins is empty: they are produced
constitutively and stochastically by the gates. We assume that the
sites $a,b,c$ are associated with the same reaction rate $r$, resulting
in a model with four parameters $r,\varepsilon,\eta,\delta$. Furthermore,
since the dynamics of the network depends only on the relative rates
of these parameters, we can arbitrarily fix the value of one parameter
in order to study the effects of the other three. Here we fix the
constitutive rate of protein production $\varepsilon$ at a nominal
value of 0.1, and vary the rates of protein degradation $\delta$,
gene unblocking $\eta$ and gene repression $r$. The results of the
parameter variation are shown in Figure 3.

Figure 3~(i) shows the
simulation results for $\eta=0.00001$, $\delta=0.001$ and $r=1.0$.
We observe alternate cycles of protein production, where each cycle
is characterised by a dominant protein. The cycles alternate in a
specific sequence of proteins $P(c)$, $P(b)$, $P(a)$ and the population
of the dominant protein stabilises at about 100 in each cycle. The
dominant population fluctuates significantly due to stochastic noise
in the system, and the duration of the cycles also varies considerably.
We can improve on all these aspects of the repressilator model by
adjusting its parameters appropriately. 

First, we observe that the dominant protein population stabilises
at an equilibrium between production and degradation, given by $\varepsilon/\delta$.
We can limit the relative size of the fluctuations by decreasing the
degradation rate to $\delta=0.0001$, resulting in a dominant population
of about 1000, as shown in Figure 3~(ii).

Next, we observe that when one protein is dominant the other two proteins
are absent and their corresponding genes are blocked, where one of
the blocked genes is actively repressed. If the repressed gene unblocks
then it is immediately blocked again by the dominant protein. If the
unrepressed gene unblocks then it can start to produce proteins, which
will repress the dominant gene and will themselves become dominant.
The duration of protein cycles is highly irregular, since it depends
on the rate of unblocking of the unrepressed gene, which is characterised
by an exponential distribution. Furthermore, both blocked genes are
in a stochastic race to unblock, and the duration of protein cycles
will also depend on how far apart they unblock from one another, which
is highly variable. We can reduce this variability by increasing the
rate of gene unblocking to $\eta=0.0001$. As this rate is increased,
the effect of degradation plays a role in improving the regularity
of oscillations: if a gene unblocks, it is immediately blocked again
by any repressors that have not yet degraded. As a result, a gene
can only start producing proteins when all residual repressors are
degraded. Since the decay curve of each protein is fairly regular,
we observe an increased regularity in the oscillations. In this setting,
a gene can repeatedly block and unblock many times while waiting for
the residual repressors to degrade. Unfortunately, this also increases
the likelihood of a leaky production of proteins, which results in
a \emph{stuttering} of the oscillations, as observed in Figure 3~(iii). 

We can compensate for this by increasing the rate of gene repression
to $r=10.0$. In this setting, even if there is one protein remaining,
it will still have a high probability of blocking the corresponding
gene. This significantly reduces the probability of a leaky production
of proteins, thereby reducing the stuttering effects, as shown in
Figure 3~(iv). 

We summarise the results of our parameter analysis for the repressilator
network:

\begin{itemize}
\item The rate of protein degradation $\delta$ should be low enough so
that the population of the dominant protein is large relative to its
fluctuations. 
\item The rate of gene unblocking $\eta$ should be higher than the rate
of protein degradation, to enable protein cycles of regular duration.
\item The rate of gene repression $r$ should be high enough that a single
protein will cause the gene to block before transcription can occur,
to prevent the leaky production of proteins. The rate should also
take into account the number of times that a gene can attempt to produce
before the last repressor has degraded, which is determined by $\eta/\delta$.
\end{itemize}
Using these basic principles we can design effective repressilator
networks with a wide range of parameters. In particular, successful
designs should include all models that satisfy the constraints $\delta<\varepsilon/1000$,
$\eta>\delta$ and $r>100\cdot\varepsilon\cdot\eta/\delta$. Additional
details are provided in the online supplementary material.

We also note that the behaviour of the stochastic repressilator significantly
differs from its deterministic counterpart. In Section IV we provide
the derivation of the ode system that follows from the kinetic reaction
scheme. While the stochastic repressilator oscillates readily without
cooperativity, it can be shown that this is not the case for the deterministic
dynamics.

\subsection{Transcription, Translation and Repressor Binding}

The repressilator network in the previous section was constructed
using a highly simplistic model of a gene gate. In this section we
examine various refinements to our gene gate model, and test whether
the results of our parameter analysis are still applicable. Note that
the high-level definition of the repressilator network remains unchanged
as $g(a,b)\mid g(b,c)\mid g(c,a)$. We simply refine our model of
a gene gate to include more biological details. 

Figure 4 presents a model of a gene
gate which considers gene transcription and RNA translation. The
simulation results with $\delta=\eta=0.0001$, $r=10$ and $\varepsilon2=\delta2=0.01$
are almost identical to those of Figure 3~(iv),
suggesting that our parameter analysis is still applicable. Here we
fix $\varepsilon/\delta2=10$ so that there is a continuous supply
of a few RNA molecules to enable steady translation, and we fix $\varepsilon2/\delta=100$
so that the dominant protein population stabilises at about 1000. 

Figure 5~(i) presents a model of a
gene gate in which a repressor must remain bound in order to block
the gene. In this situation the simulation results do not produce
alternating protein cycles when the rate of repressor binding is high,
as shown in Figure 6~(i).
This is because, when a repressor unbinds from a gene it has a high
probability of re-binding, which gives rise to a situation in which
all three genes are blocked. However, we do get oscillations when
the rate of repressor binding is very low ($r=0.00001$) as shown
in Figure 6~(ii), though
the cycles are irregular. The low repression rate ensures that a single
repressor has a low probability of switching off a gene. This allows
the gene to produce proteins when the repressor finally does unbind,
in order to start the next cycle. This also means that a large number
of repressors is required in order for a gene to be switched off.
We observe that a gene is typically switched off after about 100 repressors
are produced. In this model it is also important for the DNA-TF complex
to be long-lived ($\eta=0.00001$) so that the repressor remains tightly
bound for a sufficient length of time, comparable to the duration
of a protein cycle. Unfortunately, low $\eta$ also means that the
oscillations do not occur at regular intervals, since the duration
of protein cycles is determined by $\eta$ as opposed to the smooth
repressor degradation curve. If we increase $\eta$ to $0.0001$ we
no longer obtain distinct oscillations, since the repressor can unbind
too soon, after which the gene has a much higher probability of producing
a protein than becoming blocked again. This causes the protein cycles
to interfere with each other, as shown in Figure 6~(iii). 

Interestingly, we can solve this problem by allowing proteins to degrade
when still bound to a gene. We model this by replacing the definition
of $P'$ with $P'(b,u)=?u.P(b)+?u$ in Figure 5~(i),
which is equivalent to adding a reaction $g'\rightarrow_{\eta}g$.
This produces the desired oscillations, shown in Figure 6~(iv).
At first glance the degradation of bound repressors may seem counterintuitive,
but it can also be viewed as an abstraction of a more general requirement,
which is that a repressor can somehow dissociate from a DNA binding
site in an inactive form, such that it has very low probability of
re-binding. One way of achieving this is to allow two repressors to
bind to the DNA, as shown in Figure 5~(ii).
For large repressor populations, when a repressor unbinds it is more
likely to bind again than for the second repressor to unbind. Conversely,
for small repressor populations when a repressor unbinds it is much
less likely to bind again. In this way, the population of repressors
can be used to control the likelihood of gene activation, giving rise
to more regular cycles. Corresponding simulation results are shown
in Figure 6~(v). Although
the protein cycles are still noisy, they are nevertheless of reasonably
similar duration. 

We summarise the results for our more detailed repressilator models:

\begin{itemize}
\item The presence of gene transcription and RNA translation does not significantly
perturb the dynamics of the repressilator network, provided there
is a continuous supply of a few RNA molecules. 
\item If we assume that a protein must remain bound in order to repress
a gene then we can still obtain the desired repressilator dynamics,
provided the bound proteins can also degrade. The degradation of bound
repressors is not essential for oscillations, but it does produce
a significant improvement in their regularity.
\end{itemize}

\subsection{Cooperativity by Repressor Dimerization and Tetramerization}

As a final modification of the stochastic repressilator we discuss
the effect of cooperativity in transcription factor binding. For this
we address the cases of dimerization and tetramerization. The gate
reaction kinetics and the stochastic $\pi$-calculus models for these
two cases are depicted in Figure 7.

In the first model the gene produces a protein that can form a dimer
by sending or receiving on site b2, and the resulting dimer can send
on site b. In the second model, the dimer can form a tetramer by sending
or receiving on site b4, and the resulting tetramer can send on site
b. This way of modeling dimerization is also compatible with biological
reality, since a protein must be able to interact both on a site and
on its complement in order to dimerize. 

Figure 8 shows the
effect of cooperativity on the repressilator network. The results
on the left correspond to the repressilator with no cooperativity,
while the results on the right correspond to the repressilator with
tetramerization. The program code for the simulations is given in
supplementary online material. Figure 8~(i)
shows the populations of the three proteins $P(a)$, $P(b)$, $P(c)$
over time. We observe that the populations fluctuate significantly
less in presence of cooperativity. We can quantify this by measuring
the variability of the dominant protein populations over time. In
order to obtain a clean separation of protein cycles, we only consider
the dominant population of a given protein when the remaining two
proteins are \emph{off}. The principle of the approach is illustrated
in Figure 8~(ii).
We assume that a protein is \emph{on} when its population is above
a certain threshold, and \emph{off} when its population is below this
threshold, and we fix the threshold at roughly 10\% of the observed
steady state of protein levels, i.e. at about 100. We use this definition
to extract the dominant protein populations from the simulation results
in row (i) by application of a simple filter, in order to obtain the
plots in row (ii). The gaps in the plots correspond to situations
where multiple proteins are \emph{on} simultaneously, which we deliberately
ignore. This is a convenient metric for comparing the variability
of dominant protein populations, since it filters out situations where
multiple proteins have competing populations. We quantify the difference
between the two models by measuring the mean and standard deviation
of the dominant protein populations over a time period of $10^{7}$
time units. In absence of cooperativity we observe a mean of 880 and
a standard deviation of 196, whereas with tetramerization we observe
a mean of 935 and a standard deviation of 98. For clarity, only the
first $10^{6}$ units are shown in Figure 8~(i,ii,iii).
For a more coarse-grained comparison over the same time period, in
absence of cooperativity we observe that the dominant protein population
falls below a threshold of 800 roughly 23\% of the time, whereas with
tetramerization it falls below this threshold only 3\% of the time.
In this setting, cooperativity acts to improve the regularity of oscillations
by reducing the fluctuations in protein levels. In presence of cooperativity,
the leaky transcription of a gene is less likely to perturb the oscillations,
since at least two proteins must be produced in order to have an effect
in the case of dimerization, and at least four proteins are required
in the case of tetramerization. Thus, cooperativity can be seen not
as an essential requirement for oscillations, but as a means of improving
the stability of oscillations over a wider range of parameters.

We can compare the regularity of oscillations by measuring the duration
of protein cycles. Here we assume that a protein cycle starts when
the protein is switched \emph{on}, and ends when the next cycle starts.
Figure 8~(iii) shows
histograms of the duration of protein cycles for the three proteins.
For both models there are approximately 140 protein cycles, and we
observe a moderate improvement in cycle regularity in presence of
cooperativity. Without cooperativity we observe a mean duration of
69000 and a standard deviation of 16000, and with tetramerization
we observe a mean duration of 75000 and a standard deviation of 14000.

Note that in presence of cooperativity the rate of gene repression
$r$ can be significantly lower than in absence of cooperativity,
while still observing regular protein cycles. Not only does this improve
the robustness of the network by allowing for a broader range of parameters,
it could also be important in situations where the rate of repression
is limited by cellular constraints. For example, if we assume that
the rate of protein-gene interaction is determined by random diffusion,
it may be physically impossible for this rate to be above a certain
threshold. Cooperativity could be one way for a cell to overcome this
limitation.

\section{\label{sec:The rate equations of the gene gates}The rate equations of the gene gates}

For completeness we establish how the gate reaction kinetics can be expressed
in terms of rate equations (ode's) by making use of the mass action law.

We demonstrate this by applying the scheme to the simplest circuit that can be built from
the inhibitory gate, the autoinhibitory loop (Fall C P et al, 2002), 
where the output $B$ acts upon its own gate, hence $B$ and $A$ have to be 
identified; we first ignore the formation of {\it protein} complexes. 

With the identification $ A = B $ in the gate reaction kinetics in Figure 1, 
the autoinhibitory loop is given by 
\begin{equation}   \label{al1}
A + g \rightarrow_r g' + A\, ,
\end{equation}
\begin{equation}   \label{al2}
g \rightarrow_{\varepsilon} g + A \, ,
\end{equation}
\begin{equation}  \label{al3}
g' \rightarrow_{\eta} g\,,
\end{equation}
\begin{equation}  \label{al4}
A \rightarrow_{\delta} 0\, .
\end{equation}
In order to have a well-defined  continuous setting we consider a cellular 
environment with a protein concentration $[A]_c$ (mol/L).  We choose a
population of autoinhibitory loops with a concentration $[N]_c$ (mol/L) and
normalize according to
\begin{equation}
[A] \equiv \frac{[A]_c}{[N]_c}\,,\,\,\, [g] \equiv \frac{[g]_c}{[N]_c}\,,\,\,\, [g'] \equiv \frac{[g']_c}{[N]_c}
\end{equation}
so that $[A], [g], [g'] $ are concentration ratios, hence dimensionless quantities. 
For the gate states we have the conservation condition
 $[g] + [g'] = 1$. 
Casting the reaction kinetics into ordinary differential equations we have
\begin{eqnarray}
\dot{[A]} & = & \varepsilon [g] - \delta [A]\,, \label{oda} \\
& & \nonumber \\
\dot{[g]} & = & - r[g][A] + \eta [g']\, ,\\
& & \nonumber\\
\dot{[g']} & = & r[g][A] - \eta [g']\, .
\end{eqnarray}
Note that due to our choice of dimensionless variables the kinetic parameters
carry the same dimensions as the rates in the reaction scheme ($s^{-1}$) and we can
therefore leave the same symbols.
Using the conservation condition we can eliminate the equation for
$[g']$ and end up with only one equation for the unblocked gate state $[g]$, i.e.,
\begin{equation}  \label{odg}
\dot{[g]} = \eta(1 - (1 + \nu [A])[g])\,,
\end{equation}
where $\nu \equiv r/\eta $. 
The inhibitory loop is therefore described by two ode's, one each for 
$[A]$ and $[g]$. 

We can relate the gene gate description to the common continuum 
description of the dynamics of gene networks. This can be achieved by 
making some additional simplifications which are of 
approximate nature. First, we observe that in the limit   
$\eta \rightarrow 0 $, when $\nu \equiv r/\eta$ is kept finite,
$\dot{[g]}$ can be made very small without affecting the equation for $[A]$ 
since it does not depend on $\eta$. Therefore, this limit allows to separate 
the timescales of the dynamics of $[A]$ and $[g]$. For $\dot{[g]} \approx 0 $, 
$[g]$ varies according to
\begin{equation} \label{gA}
[g](t) \approx \frac{1}{1 + \nu [A](t)}\, .
\end{equation}
Inserting this equation into eq.(\ref{oda}) one obtains an 
equation for $[A]$ which is given by
\begin{equation} \label{hill}
\dot{[A]} = \frac{\varepsilon}{1 + \nu [A]} - \delta [A]\, .
\end{equation}
This is the standard Hill-type equation for an inhibitory loop 
(Cherry J L $\&$ Adler F R, 2000, Fall C P et al, 2002) in the case of a non-cooperative inhibition.

We can easily check the quality of this approximation. From eq.(\ref{gA}) 
we obtain
\begin{equation} \label{gapprox}
\dot{[g]} = - \frac{\nu \dot{[A]}}{(1 + \nu [A])^2} = 
- \frac{\nu}{(1 + \nu [A])^2}\left(\frac{\varepsilon}{1 + \nu [A]} 
- \delta [A]\right)  
\end{equation}
which shows that $\dot{[g]} = 0 $ is strictly fulfilled only at the stationary 
points of the dynamics of $[A]$. Due to the $[A]$-dependence of the denominator 
in the equation the time-variation of $[g]$ becomes indeed small if $[A]$ is large; 
but for small values of $[A]$, and away from the stationary
points, the approximation becomes increasingly poor. For the
autoinhibitory loop it can indeed be seen from the numerical solutions of the
equations that for large initial values of
$[A]$, and near the stationary state for $t \rightarrow \infty$, the exact and 
approximate solutions
coincide, but for the intermediate range of concentrations, both do differ
quantitatively (not shown).

For the fixed-points of the full system we find from $\dot{A} = \dot{g} = 0$ the conditions
\begin{equation}
\varepsilon g_0 = \delta A_0\,\,,\,\,\,\, g_0 = \frac{1}{1 + \nu A_0}
\end{equation}
which lead to a unique equilibrium solution. Perturbations around the 
fixed-point value $(A_0,g_0)$ in the form $ A = A_0 + \delta A $
obey 
\begin{eqnarray}
\left(
\begin{array}{c}
\delta \dot{[A]} \\
\delta \dot{[g]} 
\end{array}
\right) 
& = & 
\left(
\begin{array}{cc}
-\delta 					&  \varepsilon        \\
-\frac{\eta\nu}{1 +  \nu A_0}       &  -\eta(1 + \nu A_0) 
\end{array}
\right) 
\cdot 
\left(
\begin{array}{c}
\delta [A] \\
\delta [g] 
\end{array}
\right) 
\end{eqnarray}
Denoting the matrix in this equation by ${\bf J}$, one has $Det\, {\bf J} > 0 $ and
$Tr\, {\bf J} < 0 $ independent of the parameter values, hence the equilibrium 
state is indeed stable (Fall C P et al, 2002). 
\\

We now allow for dimerization of the transcription factor $A$. In the gate reaction kinetics
we have 
\begin{equation}
A + A \rightarrow_{d} A2\, .
\end{equation}
and the dimers degrade according to
\begin{equation}
A2 \rightarrow_{\delta} 0\, .
\end{equation}
We now assume that the dimers activate the gene according to (see Figure 6, top)
\begin{equation}
A2 + g \rightarrow_r  g' + A2\, .
\end{equation}
The rate equation for the dimers thus reads as
\begin{equation}
\dot{[A2]} = d [A]^2 - \delta [A2]\, .
\end{equation}
Upon assuming that the dimerization reaction is in equilibrium, $\dot{[A2]} = 0$,
we can relate the concentrations of dimers, $[A2] $, to $[A]^2 $, and define an
equilibrium constant $K_D$.  This leads to a modification of eq.(\ref{odg})
\begin{equation}  \label{odg2}
\dot{[g]} = \eta(1 - (1 +   \nu_D [A]^2)[g])\,,
\end{equation}
with $\nu_D \equiv \nu K_D$. The denominator in equation (\ref{hill}) is then replaced by 
\begin{equation}
\frac{\varepsilon}{1 + \nu [A]} \rightarrow \frac{\varepsilon}{1 + \nu_D [A]^2}
\end{equation}
where the exponent $h = 2 $ is the Hill-exponent corresponding to dimerization.
The case of tetramerization can be treated analogously. 

We finish this section by writing down the ode-equations of the repressilator. 
It consists of a three-gene circuit in which each gene 
represses the transcription of one of the other genes in a circular manner, e.g.
$ g_1 \dashv g_2 \dashv g_3 \dashv g_1 $.

For the deterministic version of the repressilator
the gene gate equations read as follows. 
Denoting the corresponding transcription factors of the 
repressilator genes by $[A_i]$, with $ i = 1, 2, 3 $, the rate equations
of the repressilator are given by the six ODE's 
\begin{equation} \label{a1}
\dot{[A_i]} = \varepsilon [g_{i}] - \delta [A_i]
\end{equation}
\begin{equation} \label{a2}
\dot{[g_i]}= \eta(1 - (1 + \nu_h [A_{i-1}]^h) [g_i])
\end{equation} 
with periodic conditions on the indices ($[g_4] \equiv [g_1] $). 

In equations (\ref{a1}), (\ref{a2}), the gene-transcription factor interaction
is assumed to be cooperative with a general Hill exponent $h$ whose
value is left unspecified here; dimerization corresponds to the value $h=2$
and tetramerization to $h = 4$. For a deterministic version of the repressilator, 
called `RepLeaky' by the authors, it has recently been shown that  a sufficient 
criterion on the Hill exponent is $ h > 4/3 $ in order to bring about sustained 
oscillations (M\"uller S et al, 2006). Although the RepLeaky-repressilator is 
formulated in terms of a protein-mRNA model, and hence its nonlinear 
dynamic equations thus differ from those of our gate-based version,
it turns out that the result by M\"uller et al. also applies to our case. 
This follows from a comparison of the stability analysis of both models,
which shows that the equations governing the linear stability of the fixed-points 
can be mapped onto each other. 
Hence, according to the sufficient criterion for oscillations developed by 
M\"uller et al, the non-cooperative repressilator which we found to oscillate
readily in its stochastic case, does not oscillate in its deterministic version
since $h = 1$.

Finally, it is instructive to compare the ode-description of the repressilator based 
on the gene gates with other ode-description of gene regulatory circuits,
see e.g. (Widder S et al, 2007). Here, the modeling of
simple gene circuits starts with considerably more biological detail
than our gene gate description, which is minimalistic. However, keeping
all the details is often difficult if not impossible, and sometimes not even
needed. We believe it is therefore more reasonable to start with a basic
model and do refinements at a later stage. 
At the lowest level of detail, the basic model could ultimately be simplified to the level of stochastic boolean networks, by ignoring the protein species and modeling the interaction between gene gates directly.
This is clearly difficult in modeling
schemes that are not sufficiently modular, one clear advantage of the
stochastic $\pi$-calculus, as presented in this work.

\section{Conclusion: Contributions and Relation to other Work}

To conclude, we have presented an approach to model gene regulatory 
networks which is fully compositional and stochastic due to the use of a process
calculus description of gene gates. It is made possible by exploiting
the compositional features of the stochastic $\pi$-calculus, which greatly 
facilitates the exploration of model design through simulation. 

Our approach demonstrates that for a better 
understanding of the effect of regulatory mechanisms, a coarse model can indeed be 
a useful starting point; we have shown how a stepwise modification of such a model 
can provide novel insights into their role. For our system at hand, the repressilator, we could
establish that stochasticity alone is sufficient to bring about oscillatory behaviour in the
three-gene network, and that, contrary to the deterministic case, cooperative mechanisms
are not needed. The latter are not without effect, as we could see: cooperativity of binding 
regularises the oscillations. Furthermore, we have shown how additional mechanisms
in transcription, translation and repressor binding influence the oscillatory behaviour of
the network.

We like to note that the non-cooperative case is also not entirely academic: while in 
protein binding cooperativity is mostly present in in-vivo cellular genetic networks, 
it now becomes technically feasible to 
build artificial transcriptional oscillators based on DNA and RNA which lack 
protein binding cooperativity in the transcriptional initiation process 
(Simmel F, 2007, Kim J et al., 2006). 
The properties of such artificial and in-vitro networks can thus be expected to
yield novel information about the functional constraints of gene regulation systems,
in particular when combined with mathematical modeling.

We have also established the relationship between the stochastic and the 
deterministic (ode) description of the gene gates. 
While both descriptions are valid representations 
of the underlying gate reaction kinetics, the example of
the repressilator clearly shows that both descriptions do not yield equivalent 
system behaviour.

Stochastic effects in networks have previously been studied mostly 
in the context of their role in perturbing an underlying deterministic 
dynamics. Also there surprising effects were observed, like the occurrence
of oscillatory behaviour at a finite distance from a Hopf bifurcation,
or even oscillations via a different type of bifurcation 
(Freidlin M I, 2001; Lee Deville R E et al, 2006). We stress that in our context 
stochastic effects do not act merely as perturbations of an underlying deterministic 
dynamics, but bring about the dynamic behaviour in the first place. 

It remains a challenge to find the correct abstraction level for the representation of 
the biologically relevant features of a regulatory network in terms of computable elements.
In this respect our compositional approach is of advantage, since it permits to modify the 
properties of the individual components by fine tuning without affecting the overall 
network topology.  
\\

{\bf Acknowledgement.} RB wishes to thank the Microsoft Research - University
of Trento Centre of Computational Systems Biology for its hospitality 
during a stay, and Fran{\c c}ois Fages, Leon Glass and Marc Lefranc for 
illuminating discussions. 
\\

\newpage
%%%\input{references_rev}

%\newpage
\clearpage

\begin{figure}[h]
\includegraphics[scale=0.8]{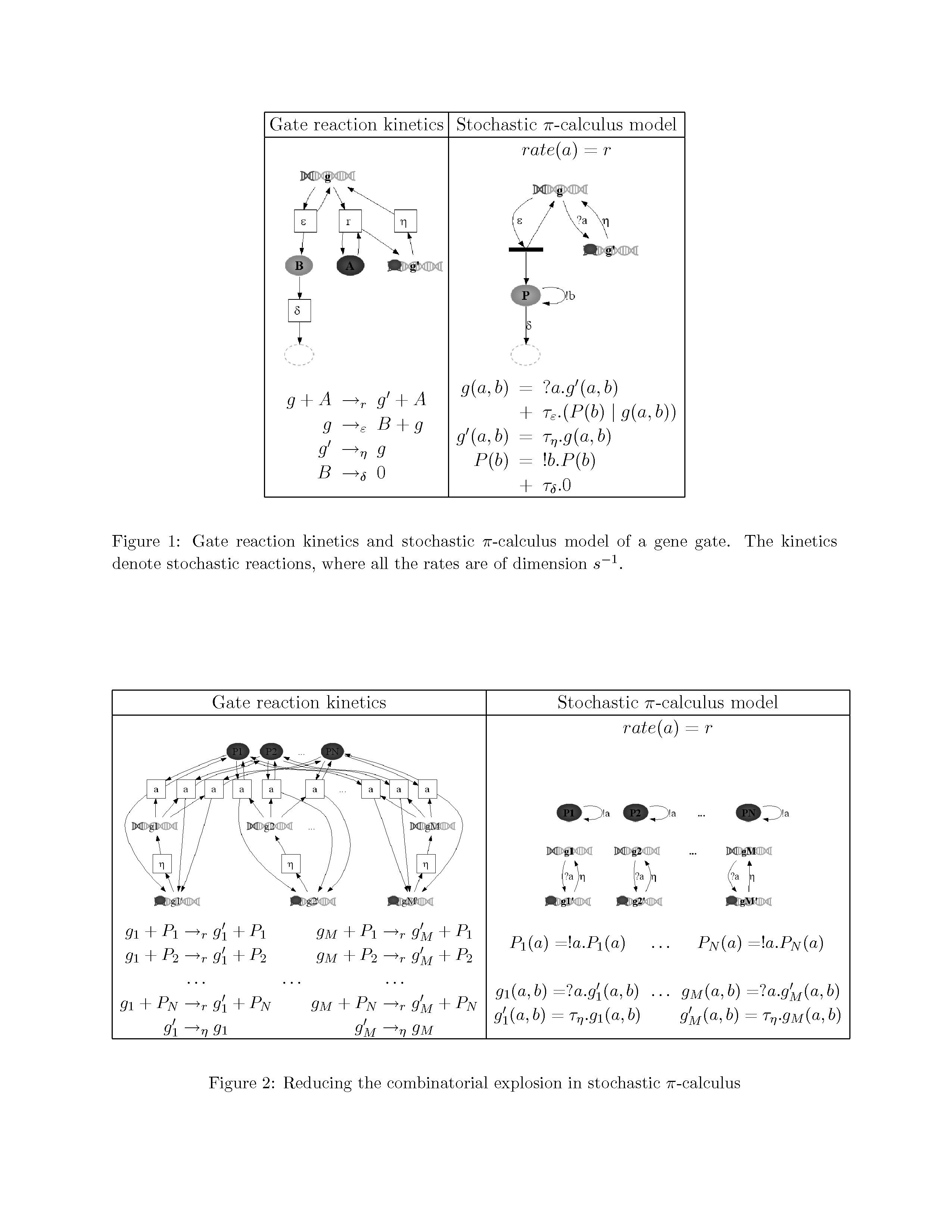}
\end{figure}
%
%\newpage
\clearpage

\begin{figure}[h]
\includegraphics[scale=0.8]{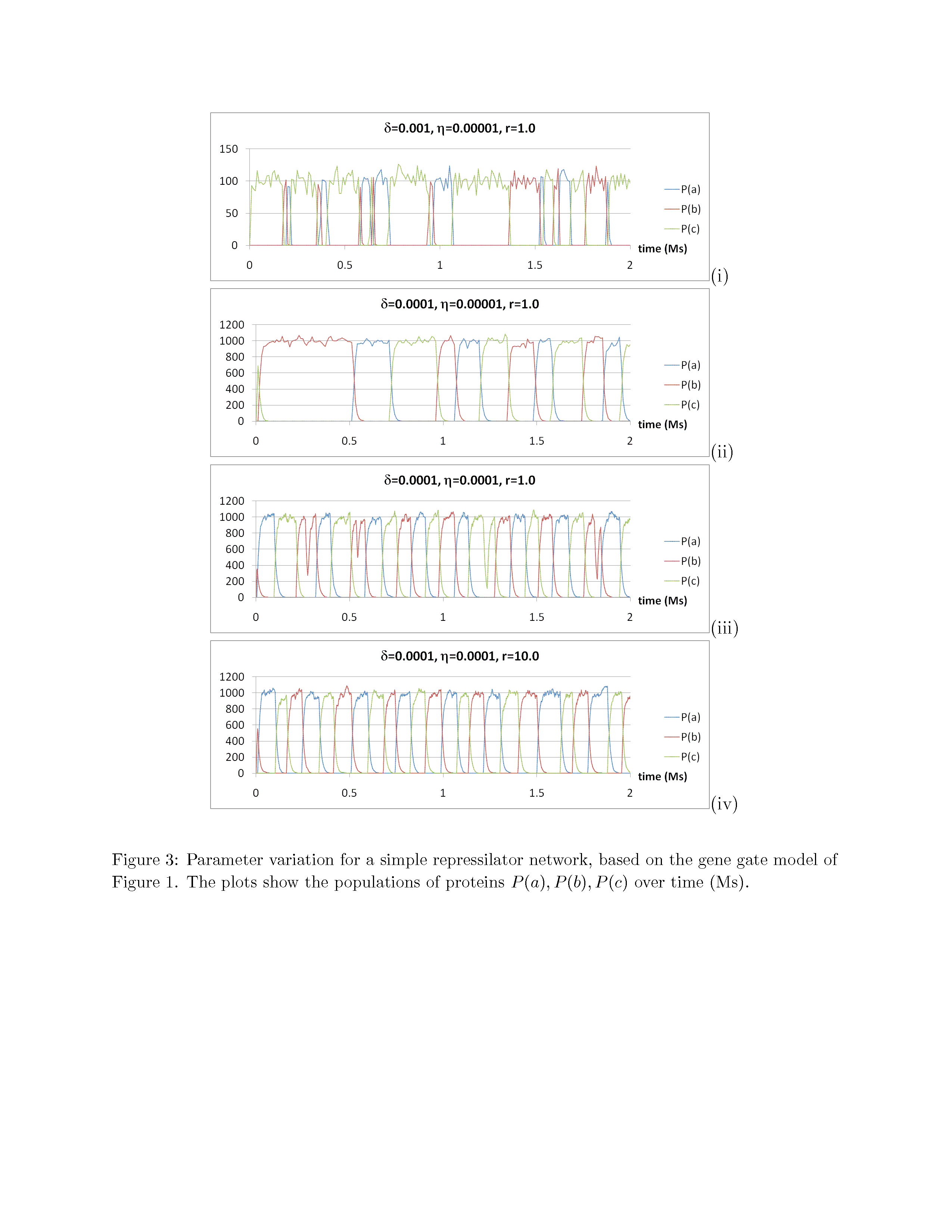}
\end{figure}
%
%\newpage
\clearpage

\begin{figure}[h]
\includegraphics[scale=0.8]{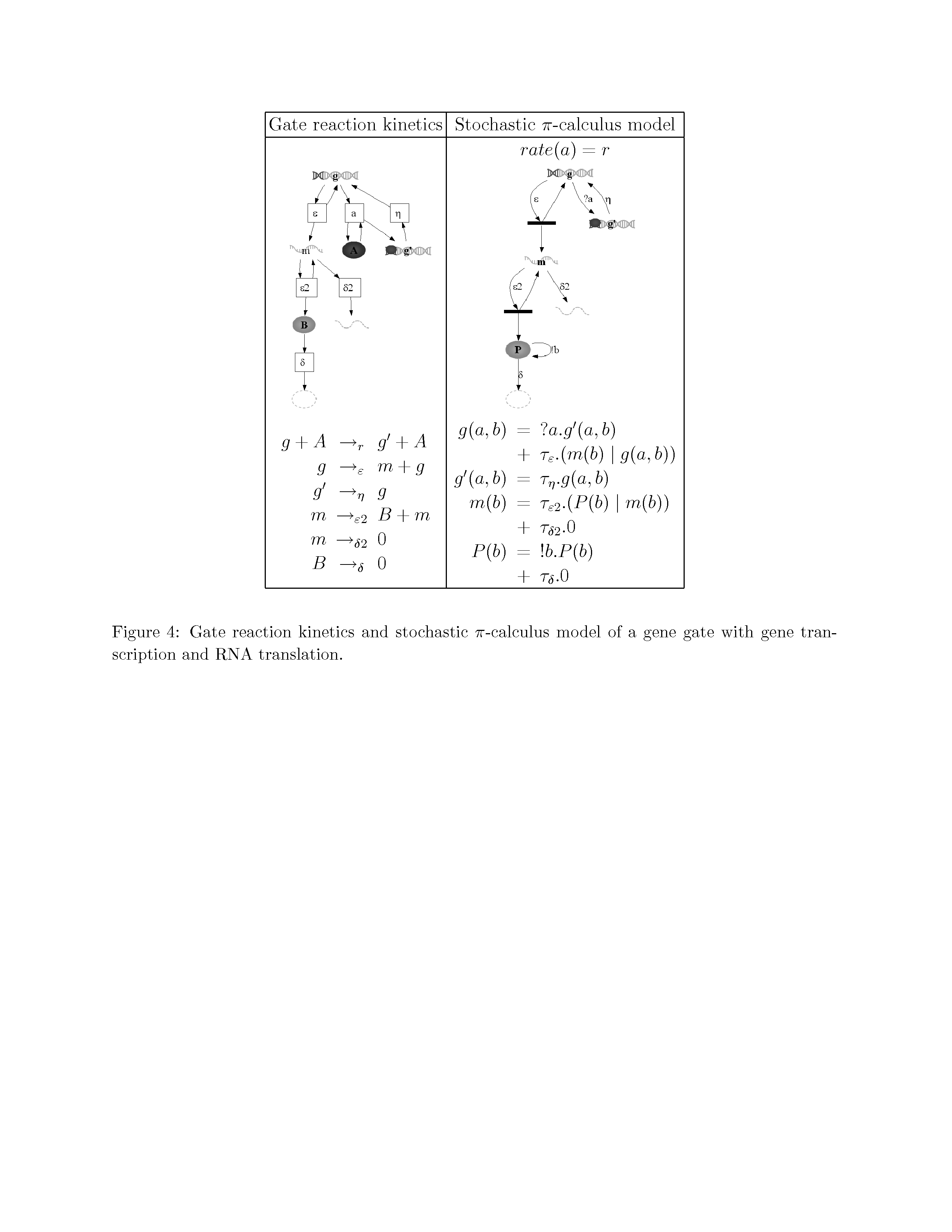}
\end{figure}
%
%\newpage
\clearpage

\begin{figure}[h]
\includegraphics[scale=0.8]{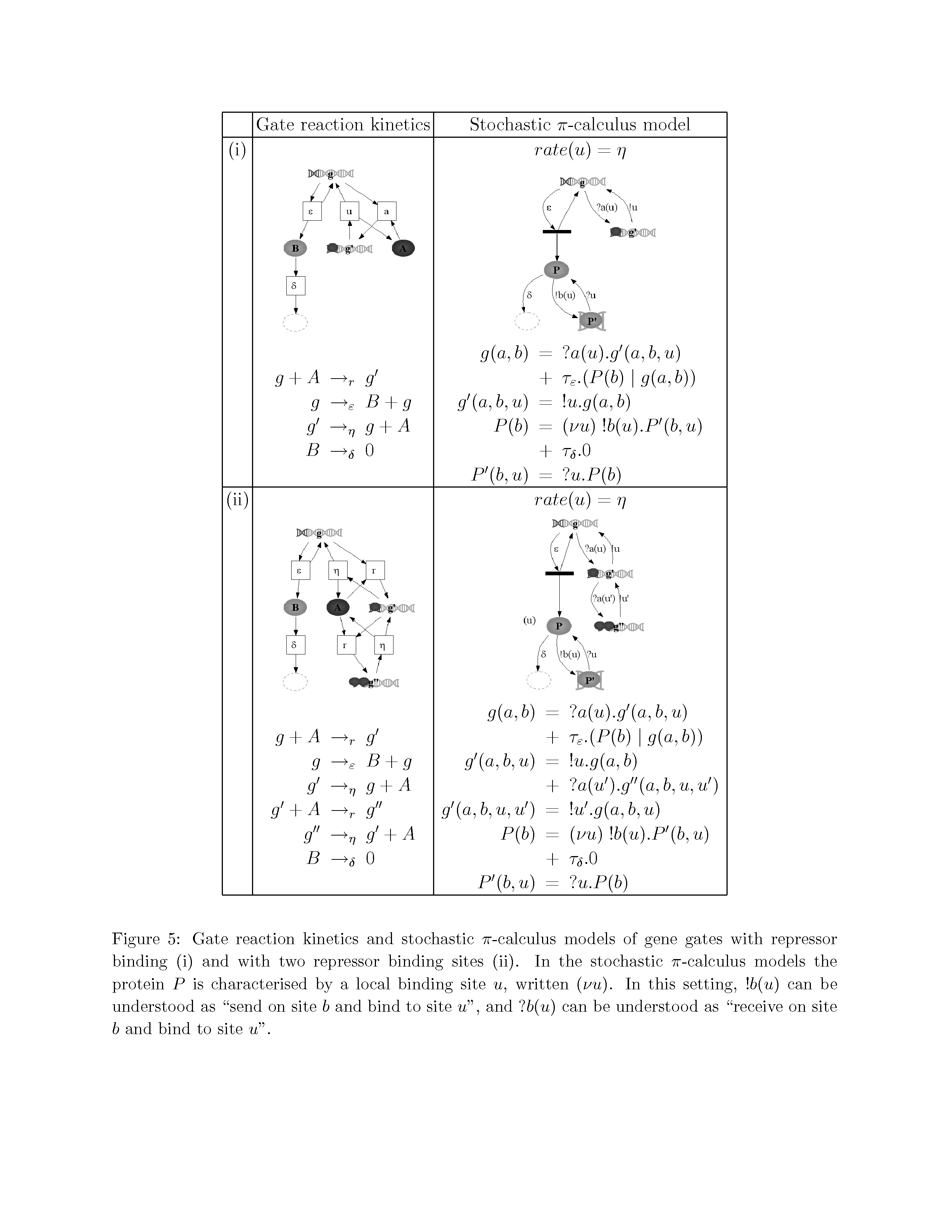}
\end{figure}
%
%\newpage
\clearpage

\begin{figure}[h]
\includegraphics[scale=0.8]{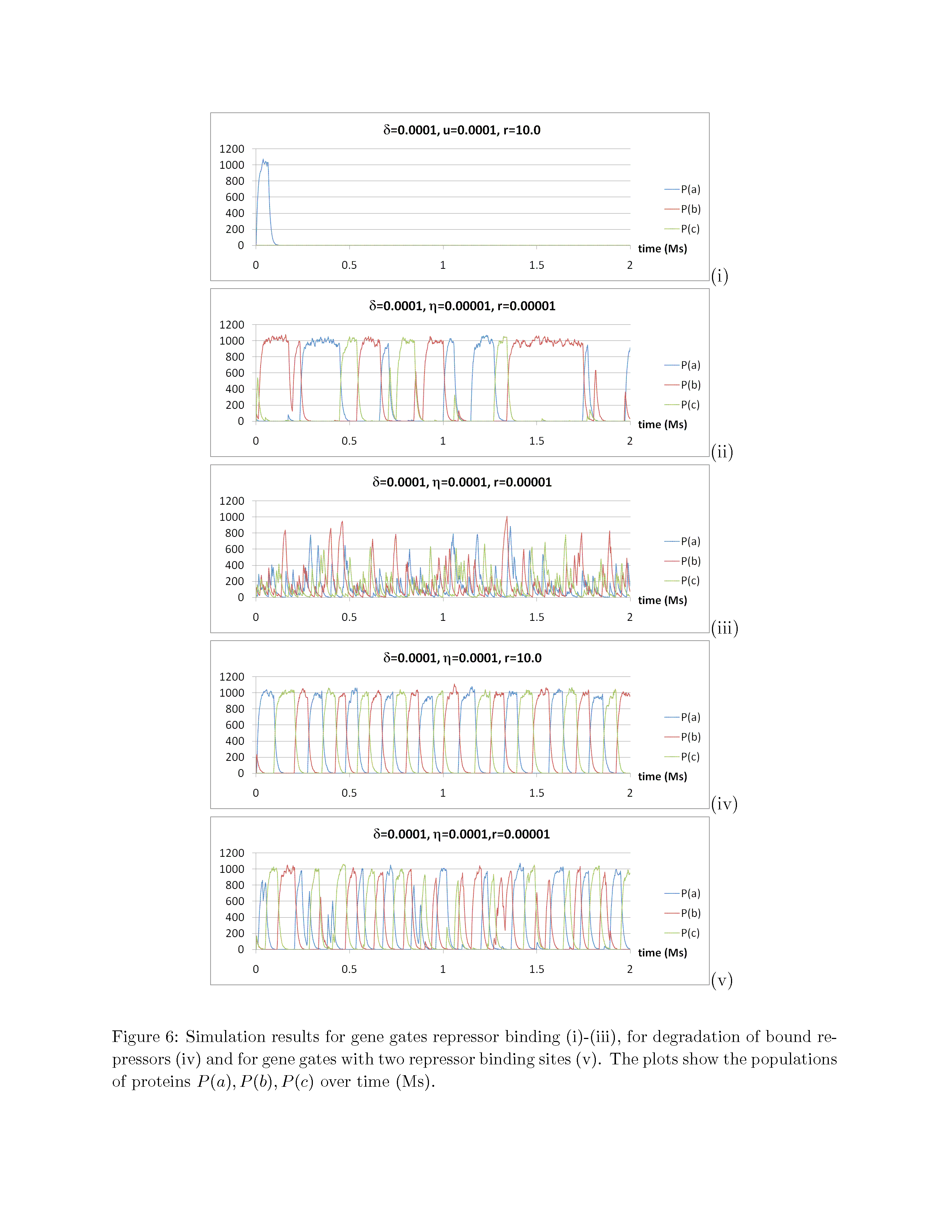}
\end{figure}
%
%\newpage
\clearpage

\begin{figure}[h]
\includegraphics[scale=0.8]{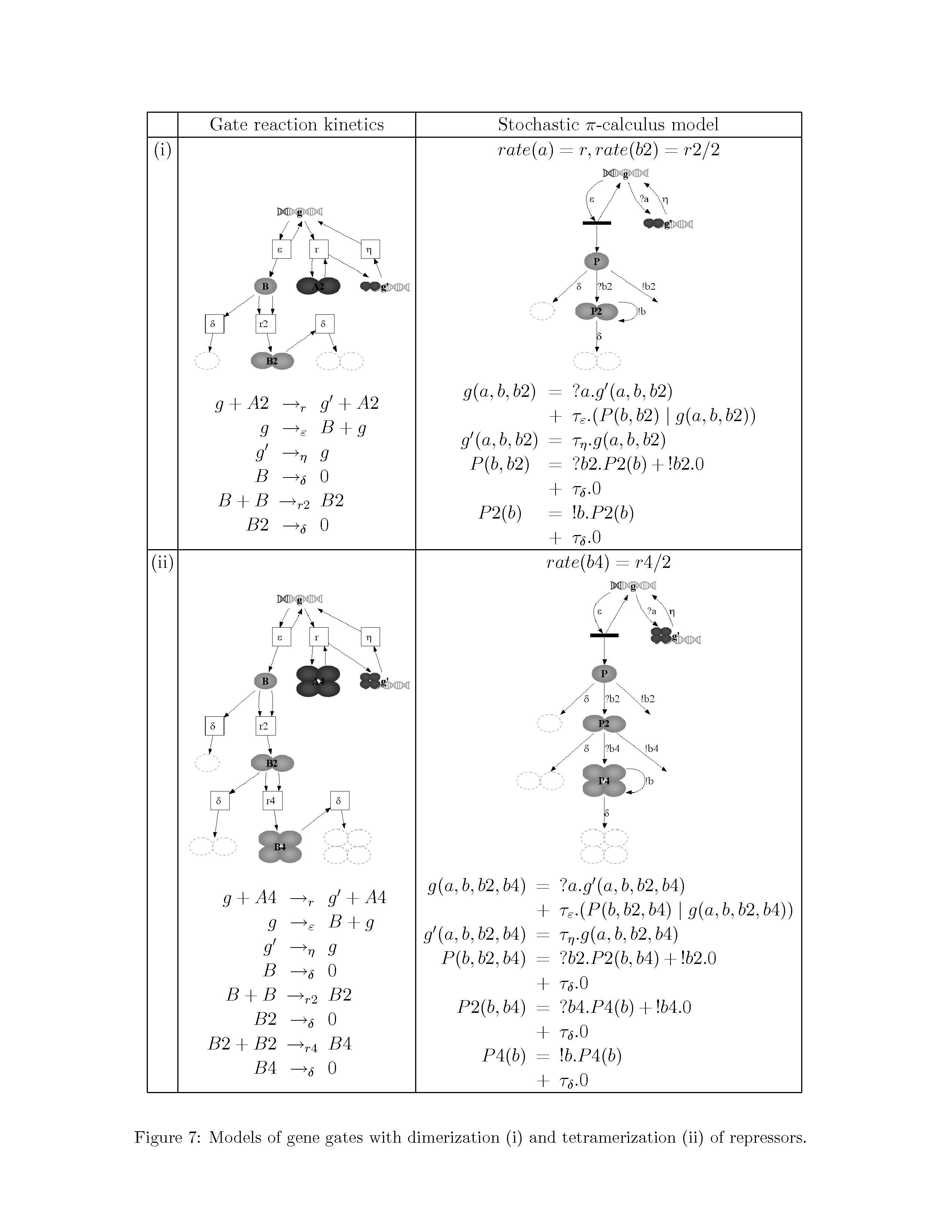}
\end{figure}
%
%\newpage
\clearpage

\begin{figure}[h]
\includegraphics[scale=0.8]{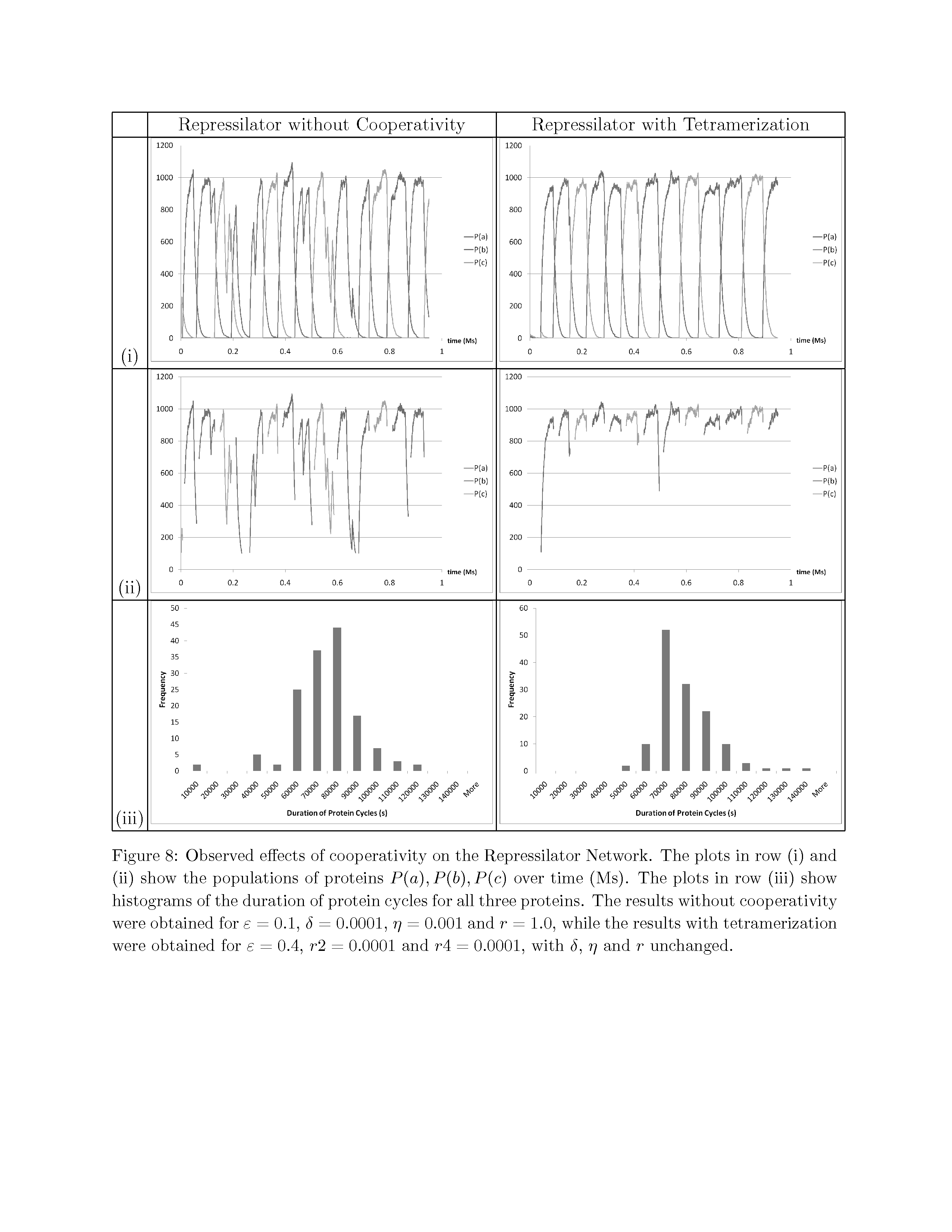}
\end{figure}
\end{document}